# Complexity Results about Nash Equilibria

Vincent Conitzer        Tuomas Sandholm

May 2002

CMU-CS-02-135

School of Computer Science
Carnegie Mellon University
Pittsburgh, PA 15213

## Abstract

Noncooperative game theory provides a normative framework for analyzing strategic interactions. However, for the toolbox to be operational, the solutions it defines will have to be *computed*. In this paper, we provide a single reduction that 1) demonstrates $\mathcal{NP}$-hardness of determining whether Nash equilibria with certain natural properties exist, and 2) demonstrates the #$\mathcal{P}$-hardness of counting Nash equilibria (or connected sets of Nash equilibria). We also show that 3) determining whether a pure-strategy Bayes-Nash equilibrium exists is $\mathcal{NP}$-hard, and that 4) determining whether a pure-strategy Nash equilibrium exists in a stochastic (Markov) game is $\mathcal{PSPACE}$-hard even if the game is invisible (this remains $\mathcal{NP}$-hard if the game is finite). All of our hardness results hold even if there are only two players and the game is symmetric.

This material is based upon work supported by the National Science Foundation under CAREER Award IRI-9703122, Grant IIS-9800994, ITR IIS-0081246, and ITR IIS-0121678.

# 1 Introduction

Noncooperative game theory provides a normative framework for analyzing strategic interactions. However, for the toolbox to be operational, the solutions it defines will have to be *computed* [47]. There has been growing interest in the computational complexity of natural questions in game theory. Starting at least as early as the 1970s, complexity theorists have focused on the complexity of playing particular highly structured games (usually board games, such as chess or Go [26], but also games such as Geography or QSAT [48]). These games tend to be alternating-move zero-sum games with enormous state spaces, which can nevertheless be concisely represented due to the simple rules governing the transition between states. As a result, effort on finding results for general classes of games has often focused on complex languages in which such structured games can be concisely represented.

Real-world strategic settings are generally not nearly as structured, nor do they generally possess the other properties (most notably, zero-sumness) of board games and the like. Algorithms for analyzing this more general class of games strategically are a necessary component of sophisticated agents that are to play such games. Additionally, they are needed by *mechanism designers* who have (some) control over the rules of the game and would like the outcome of the game to have certain properties, such as maximum social welfare.

Noncooperative game theory provides languages for representing large classes of strategic settings, as well as sophisticated notions of what it means to "solve" such games. The best known solution concept is that of *Nash equilibrium* [31], where the players' strategies are such that no individual player can derive any benefit from deviating from its strategy. The question of how complex it is to construct such an equilibrium has been dubbed "a most fundamental computational problem whose complexity is wide open" and "together with factoring, [...] the most important concrete open question on the boundary of $\mathcal{P}$ today" [38].

While this question remains open, important concrete advances have been made in determining the complexity of related questions. For example, 2-person zero-sum games can be solved using linear programming [27] in polynomial time. As another example, determining the existence of a joint strategy where each player gets expected payoff at least $k$ in a team game in extensive form is $\mathcal{NP}$-complete (as the number of players grows) [6]. As yet another example, in 2-player general-sum normal form games, determining the existence of Nash equilibria *with certain properties* is $\mathcal{NP}$-hard [13]. Finally, the complexity of best-responding and of guaranteeing payoffs in repeated and sequential games has been studied in [5, 20, 37, 51].

In this paper we provide new complexity results on questions related to Nash equilibria. In Section 2 we provide a single reduction which significantly improves on many of Gilboa and Zemel's results on determining the existence of Nash equilibria with certain properties. In Section 3, we use the same reduction to show that counting the number of Nash equilibria (or connected sets of Nash equilibria) is $\#\mathcal{P}$-hard. In Section 4 we show that determining whether a pure-strategy Bayes-Nash equilibrium exists is $\mathcal{NP}$-hard. Finally, in Section 5 we show that determining whether a pure-strategy Nash equilibrium exists in a stochastic (Markov) game is $\mathcal{PSPACE}$-hard even if the game is invisible (this remains $\mathcal{NP}$-hard if the game is finite). All of our hardness results hold even if there are only two players and the game is symmetric.

# 2 Equilibria with certain properties in normal form games

When one analyzes the strategic structure of a game, especialy from the viewpoint of a mechanism designer who tries to construct good rules for a game, finding a single equilibrium is far from satisfactory. More desirable equilibria may exist: in this case the game becomes more attractive, especially if one can coax the players into playing a desirable equilibrium. Also, less desirable equilibria may exist: in this case the game becomes less attractive. Before we can make a definite judgment about the quality of the game, we would like to know the answers to questions such as: What is the game's most desirable equilibrium? Is there a unique equilibrium? If not, how many equilibria are there? Algorithms that tackle these questions would be useful both to players and to the mechanism designer.

Furthermore, algorithms that answer certain existence questions may pave the way to designing algorithms that construct a Nash equilibrium. For example, if we had an algorithm that told us whether there exists any equilibrium where a certain player plays a certain strategy, this could be useful in eliminating possibilities in the search for a Nash equilibrium.

However, all the existence questions that we have investigated turn out to be $\mathcal{NP}$-hard. These are not the first results of this nature; most notably, Gilboa and Zemel provide some $\mathcal{NP}$-hardness results in the same spirit [13]. We provide a single reduction which in demonstrates (sometimes stronger versions of) most of their hardness results, and interesting new results. Additionally, as we show in Section 3, the reduction can be used to show #$\mathcal{P}$-hardness of counting the number of equilibria.

To begin, we need some standard defintions from game theory.

**Definition 1** *In a* normal form game, *we are given a set of agents $A$, and for each agent $i$, a strategy set $\Sigma_i$ and a utility function $u_i : \Sigma_1 \times \Sigma_2 \times \ldots \times \Sigma_{|A|} \to \Re$.*

**Definition 2** *A* mixed strategy *$\sigma_i$ for player $i$ is a probability distribution over $\Sigma_i$. A special case of a mixed strategy is a* pure strategy, *where all of the probability mass is on one element of $\Sigma_i$.*

**Definition 3 (Nash [31])** *Given a normal form game, a* Nash equilibrium (NE) *is vector of mixed strategies, one for each agent $i$, such that no agent has an incentive to deviate from its mixed strategy given that the others do not deviate. That is, for any $i$ and any alternative mixed strategy $\sigma'_i$, we have $E[u_i(s_1, s_2, \ldots, s_i, \ldots, s_{|A|})] \geq E[u_i(s_1, s_2, \ldots, s'_i, \ldots, s_{|A|})]$, where each $s_i$ is drawn from $\sigma_i$, and $s'_i$ is drawn from $\sigma'_i$.*

Now we are ready to present our reduction.

**Definition 4** *Let $\phi$ be a Boolean formula in conjunctive normal form. Let $V$ be its set of variables (with $|V| = n$), $L$ the set of corresponding literals (a positive and a negative one for each variable)[1], and $C$ its set of clauses. The function $v : L \to V$ gives the variable*

---

[1]Thus, if $x_1$ is a variable, $x_1$ and $-x_1$ are literals. We make a distinction between the variable $x_1$ and the literal $x_1$.

corresponding to a literal, e.g. $v(x_1) = v(-x_1) = x_1$. We define $G(\phi)$ to be the following symmetric 2-player game in normal form. Let $\Sigma \equiv \Sigma_1 = \Sigma_2 = L \cup V \cup C \cup \{f\}$. Let the utility functions be specified as follows:

- $u_1(l^1, l^2) = u_2(l^2, l^1) = 1$ for all $l^1, l^2 \in L$ with $l^1 \neq -l^2$;
- $u_1(l, -l) = u_2(-l, l) = -2$ for all $l \in L$;
- $u_1(l, x) = u_2(x, l) = -2$ for all $l \in L$, $x \in \Sigma - L$;
- $u_1(v, l) = u_2(l, v) = 2$ for all $v \in V$, $l \in L$ with $v(l) \neq v$;
- $u_1(v, l) = u_2(l, v) = 2 - n$ for all $v \in V$, $l \in L$ with $v(l) = v$;
- $u_1(v, x) = u_2(x, v) = -2$ for all $v \in V$, $x \in \Sigma - L$;
- $u_1(c, l) = u_2(l, c) = 2$ for all $c \in C$, $l \in L$ with $l \notin c$;
- $u_1(c, l) = u_2(l, c) = 2 - n$ for all $c \in C$, $l \in L$ with $l \in c$;
- $u_1(c, x) = u_2(x, c) = -2$ for all $c \in C$, $x \in \Sigma - L$;
- $u_1(f, f) = u_2(f, f) = 0$;
- $u_1(f, x) = u_2(x, f) = 1$ for all $x \in \Sigma - \{f\}$.

**Theorem 1** *If $(l_1, l_2, \ldots, l_n)$ (where $v(l_i) = x_i$) satisfies $\phi$, then there is a Nash equilibrium of $G(\phi)$ where both players play $l_i$ with probability $\frac{1}{n}$, with expected utility 1 for each player. Furthermore, the only other Nash equilibrium is the one where both players play $f$, and receive expected utility 0 each.*

**Proof**: We first demonstrate that these combinations of mixed strategies indeed do constitute Nash equilibria. If $(l_1, l_2, \ldots, l_n)$ (where $v(l_i) = x_i$) satisfies $\phi$ and the other player plays $l_i$ with probability $\frac{1}{n}$, playing one of these $l_i$ as well gives utility 1. On the other hand, playing the negation of one of these $l_i$ gives utility $\frac{1}{n}(-2) + \frac{n-1}{n}(1) < 1$. Playing some variable $v$ gives utility $\frac{1}{n}(2-n) + \frac{n-1}{n}(2) = 1$ (since one of the $l_i$ that the other player sometimes plays has $v(l_i) = v$). Playing some clause $c$ gives utility at most $\frac{1}{n}(2-n) + \frac{n-1}{n}(2) = 1$ (since one of the $l_i$ that the other player sometimes plays occurs in clause $c$, since the $l_i$ satisfy $\phi$). Finally, playing $f$ gives utility 1. It follows that playing any one of the $l_i$ that the other player sometimes plays is an optimal response, and hence that both players playing each of these $l_i$ with probability $\frac{1}{n}$ is a Nash equilibrium. Clearly, both players playing $f$ is also a Nash equilibrium since playing anything else when the other plays $f$ gives utility $-2$.

Now we demonstrate that there are no other Nash equilibria. If the other player always plays $f$, the unique best response is to also play $f$ since playing anything else will give utility $-2$. Otherwise, given a mixed strategy for the other player, consider a player's expected utility given that the other player does not play $f$. (That is, the probability distribution over the other player's strategies is proportional to the probability distribution constituted by that player's mixed strategy, except $f$ occurs with probability 0). If this expected utility is smaller than 1, the player is strictly better off playing $f$ (which gives utility 1 when the other player does not play $f$, and also performs better than the original strategy when the other player does play $f$). So this cannot happen in a Nash equilibrium.

There are no Nash equilibria where one player always plays $f$ but the other does not, so suppose both players play $f$ with probability less than one. Consider the expected social welfare ($E[u_1 + u_2]$), given that neither player plays $f$. It is easily verified that there is no outcome with social welfare greater than 2. Additionally, any outcome in which one player plays an element of $V$ or $C$ has social welfare strictly below 2. It follows that if either player ever plays an element of $V$ or $C$, the expected social welfare given that neither player plays $f$ is strictly below 2. By linearity of expectation it follows that the expected utility of at least one player is strictly below 1 given that neither player plays $f$, and by the above reasoning, this player would be strictly better off playing $f$ instead of its randomization over strategies other than $f$. It follows that no element of $V$ or $C$ is ever played in a Nash equilibrium.

So, we can assume both players only put positive probability on strategies in $L \cup \{f\}$. Then, if the other player puts positive probability on $f$, playing $f$ is a strictly better response than any element of $L$ (since both give utility 1 if the other player plays an element of $L$, but $f$ does better if the other player plays $f$). It follows that the only equilibrium where $f$ is ever played is the one where both players always play $f$.

Now we can assume that both players only put positive probability on elements of $L$. Suppose that for some $l \in L$, the probability that a given player plays either $l$ or $-l$ is less than $\frac{1}{n}$. Then the expected utility for the other player of playing $v(l)$ is strictly greater than $\frac{1}{n}(2-n) + \frac{n-1}{n}(2) = 1$, and hence this cannot be a Nash equilibrium. So we can assume that for any $l \in L$, the probability that a given player plays either $l$ or $-l$ is precisely $\frac{1}{n}$.

If there is an element of $L$ such that player 1 puts positive probability on it and player 2 on its negation, both players have expected utility less than 1 and would be better off switching to $f$. So, in a Nash equilibrium, if player 1 plays $l$ with some probability, player 2 must play $l$ with probability $\frac{1}{n}$, and thus player 1 must play $l$ with probability $\frac{1}{n}$. Thus we can assume that for each variable, exactly one of its corresponding literals is played with probability $\frac{1}{n}$ by both players. It follows that in any Nash equilibrium (besides the one where both players play $f$), literals that are sometimes played indeed correspond to an assignment to the variables.

All that is left to show is that if this assignment does not satisfy $\phi$, we do not have a Nash equilibrium. Let $c \in C$ be a clause that is not satisfied by the assignment, that is, none of its literals are ever played. Then playing $c$ would give utility 2, and both players would be better off playing this. ∎

Hence, there exists a Nash equilibrium in $G(\phi)$ where each player gets utility 1 if and only if $\phi$ is satisfiable; otherwise, the only equilibrium is the one where both players play $f$ and each of them gets 0. Since any sensible definition of welfare optimization would prefer the first kind of equilibrium, it follows that determining whether a "good" equilibrium exists is hard for any such definition. Additionally, the first kind of equilibrium is, in various senses, an optimal outcome for the game, even if the players were to cooperate, so even finding out whether such an optimal equilibrium exists is hard. The following corollaries illustrate these points (each corollary is immediate from Theorem 1).

**Corollary 1** *Even in symmetric 2-player games, it is $\mathcal{NP}$-hard to determine whether there*

*exists a NE with expected (standard) social welfare ($E[\sum_{1 \leq i \leq |A|} u_i]$) at least $k$, even when $k$ is the maximum social welfare that could be obtained in the game.*

**Corollary 2** *Even in symmetric 2-player games, it is $\mathcal{NP}$-hard to determine whether there exists a NE where all players have expected utility at least $k$, even when $k$ is the largest number such that there exists a distribution over outcomes of the game such that all players have expected utility at least $k$.*

**Corollary 3** *Even in symmetric 2-player games, it is $\mathcal{NP}$-hard to determine whether there exists a Pareto-optimal NE. (A distribution over outcomes is Pareto-optimal if there is no other distribution over outcomes such that every player has at least equal expected utility, and at least one player has strictly greater expected utility).*

**Corollary 4** *Even in symmetric 2-player games, it is $\mathcal{NP}$-hard to determine whether there exists a NE where player 1 has expected utility at least $k$.*

Some additional interesting corollaries are:

**Corollary 5** *Even in symmetric 2-player games, it is $\mathcal{NP}$-hard to determine whether there is more than one Nash equilibrium.*

**Corollary 6** *Even in symmetric 2-player games, it is $\mathcal{NP}$-hard to determine whether there is an equilibrium where player 1 sometimes plays $x \in \Sigma_1$.*

**Corollary 7** *Even in symmetric 2-player games, it is $\mathcal{NP}$-hard to determine whether there is an equilibrium where player 1 never plays $x \in \Sigma_1$.*

All of these results indicate that it is hard to obtain summary information about a game's Nash equilibria. (Corollary 5 and weaker versions of Corollaries 2, 6 and 7 were first proven by Gilboa and Zemel [13].)

## 3 Counting the number of equilibria in normal form games

Existence questions do not tell the whole story. In general, we are interested in characterizing all the equilibria of a game. One rather weak such characterization is the number of equilibria[2]. We can use Theorem 1 to show that even determining this number in a given normal form game is hard.

---
[2]The number of equilibria in normal form games has been studied both in the worst case [30] and in the average case [29].

**Corollary 8** *Even in symmetric 2-player games, counting the number of Nash equilibria is #$\mathcal{P}$-hard.*

**Proof**: The number of Nash equilibria in our game $G(\phi)$ is the number of satisfying assignments to the variables of $\phi$, plus one. Counting the number of satisfying assignments to a CNF formula is #$\mathcal{P}$-hard [49]. ∎

It is easy to construct games where there is a continuum of Nash equilibria. In such games, it would be more meaningful to ask how many distinct continuums of equilibria there are. More formally, one can ask how many maximal connected sets of equilibria a game has (a maximal connected set is a connected set which is not a proper subset of a connected set).

**Corollary 9** *Even in symmetric 2-player games, counting the number of maximal connected sets of Nash equilibria is #$\mathcal{P}$-hard.*

**Proof**: Every Nash equilibrium in $G(\phi)$ constitutes a maximal connected set by itself, so the number of maximal connected sets is the number of satisfying assignments to the variables of $\phi$, plus one. ∎

The most interesting #$\mathcal{P}$-hardness results are the ones where the corresponding existence and search questions are easy, such as counting the number of perfect bipartite matchings. In the case of Nash equilibria, the existence question is completely trivial: it has been analytically shown (by Kakutani's fixed point theorem) that a Nash equilibrium always exists [31]. The complexity of the search question remains open.

# 4   Pure-strategy Bayes-Nash equilibria

Equilibria in pure strategies are particularly desirable because they avoid the uncomfortable requirement that players randomize over strategies among which they are indifferent [12]. In normal form games with small numbers of players, it is easy to determine the existence of pure-strategy equilibria: one can simply check, for each combination of pure strategies, whether it constitutes a Nash equilibrium. However, this is not feasible in *Bayesian* games, where the players have private information about their own preferences (represented by *types*). Here, players may condition their actions on their types, so the strategy space of each player is exponential in the number of types.

In this section, we show that the question of whether a pure-strategy Bayes-Nash equilibrium exists is in fact $\mathcal{NP}$-hard even in symmetric two-player games. First, we need the standard definition of a Bayesian game and Bayes-Nash equilibrium from game theory.

**Definition 5** *In a* Bayesian game, *we are given a set of agents $A$; for each agent $i$, a set of types $\Theta_i$; a commonly known prior distribution $\phi$ over $\Theta_1 \times \Theta_2 \times \ldots \times \Theta_{|A|}$; for each agent $i$, a set of strategies $\Sigma_i$; and for each agent $i$, a utility function $u_i : \Theta_i \times \Sigma_1 \times \Sigma_2 \times \ldots \times \Sigma_{|A|} \to \Re$.*

**Definition 6 (Harsanyi [15])** *Given a Bayesian game, a* Bayes-Nash equilibrium (BNE) *is a vector of mixed strategies, one for each pair $i, \theta_i \in \Theta_i$, such that no agent has an incentive to deviate, for any of its types, given that the others do not deviate. That is, for any $i, \theta_i \in \Theta_i$, and any alternative mixed strategy $\sigma'_{i,\theta_i}$, we have*

$$E_{\theta_{-i}|\theta_i}[E[u_i(\theta_i, s_{1,\theta_1}, s_{2,\theta_2}, \ldots, s_{i,\theta_i}, \ldots, s_{|A|,\theta_{|A|}})]] \geq E_{\theta_{-i}|\theta_i}[E[u_i(\theta_i, s_{1,\theta_1}, s_{2,\theta_2}, \ldots, s'_{i,\theta_i}, \ldots, s_{|A|,\theta_{|A|}})]]$$

*where each $s_{i,\theta_i}$ is drawn from $\sigma_{i,\theta_i}$, and $s'_{i,\theta_i}$ is drawn from $\sigma'_{i,\theta_i}$.*

We can now define the computational problem that we study.

**Definition 7 (PURE-STRATEGY-BNE)** *We are given a Bayesian game. We are asked whether there exists a BNE where all the strategies $\sigma_{i,\theta_i}$ are pure.*

To show our $\mathcal{NP}$-hardness result, we will reduce from the SET-COVER problem.

**Definition 8 (SET-COVER)** *We are given a set $S = \{s_1, \ldots, s_n\}$, subsets $S_1, S_2, \ldots, S_m$ of $S$ with $\bigcup_{1 \leq i \leq m} S_i = S$, and an integer $k$. We are asked whether there exist $S_{c_1}, S_{c_2}, \ldots, S_{c_k}$ such that $\bigcup_{1 \leq i \leq k} S_{c_i} = S$.*

**Theorem 2** *PURE-STRATEGY-BNE is $\mathcal{NP}$-hard, even in symmetric 2-player games where $\phi$ is uniform.*

**Proof**: We reduce an arbitrary SET-COVER instance to the following PURE-STRATEGY-BNE instance. Let there be two players, with $\Theta \equiv \Theta_1 = \Theta_2 = \{\theta^1, \ldots, \theta^k\}$. $\phi$ is uniform. Furthermore, $\Sigma \equiv \Sigma_1 = \Sigma_2 = \{S_1, S_2, \ldots, S_m, s_1, s_2, \ldots, s_n\}$. The utility functions we choose in fact do not depend on the types, so we omit the type argument in their definitions. They are as follows:

- $u_1(S_i, S_j) = u_2(S_j, S_i) = 1$ for all $S_i$ and $S_j$;
- $u_1(S_i, s_j) = u_2(s_j, S_i) = 1$ for all $S_i$ and $s_j \notin S_i$;
- $u_1(S_i, s_j) = u_2(s_j, S_i) = 2$ for all $S_i$ and $s_j \in S_i$;
- $u_1(s_i, s_j) = u_2(s_j, s_i) = -3k$ for all $s_i$ and $s_j$;
- $u_1(s_j, S_i) = u_2(S_i, s_j) = 3$ for all $S_i$ and $s_j \notin S_i$;
- $u_1(s_j, S_i) = u_2(S_i, s_j) = -3k$ for all $S_i$ and $s_j \in S_i$.

We now show the two instances are equivalent. First suppose there exist $S_{c_1}, S_{c_2}, \ldots, S_{c_k}$ such that $\bigcup_{1 \leq i \leq k} S_{c_i} = S$. Suppose both players play as follows: when their type is $\theta_i$, they play $S_{c_i}$. We claim that this is a BNE. For suppose the other player employs this strategy. Then, because for any $s_j$, there is at least one $S_{c_i}$ such that $s_j \in S_{c_i}$, we have that the expected utility of playing $s_j$ is at most $\frac{1}{k}(-3k) + \frac{k-1}{k}3 < 0$. It follows that playing any of the $S_j$ (which gives utility 1) is optimal. So there is a pure-strategy BNE.

On the other hand, suppose that there is a pure-strategy BNE. We first observe that in no pure-strategy BNE, both players play some element of $S$ for some type: for if the other player sometimes plays some $s_j$, the utility of playing some $s_i$ is at most $\frac{1}{k}(-3k) + \frac{k-1}{k}3 < 0$, whereas playing some $S_i$ instead guarantees a utility of at least 1. So there is at least one player who never plays any element of $S$. Now suppose the other player sometimes plays some $s_j$. We know there is some $S_i$ such that $s_j \in S_i$. If the former player plays this $S_i$, this will give it a utility of at least $\frac{1}{k}2 + \frac{k-1}{k}1 = 1 + \frac{1}{k}$. Since it must do at least this well in the equilibrium, and it never plays elements of $S$, it must sometimes receive utility 2. It follows that there exist $S_a$ and $s_b \in S_a$ such that the former player sometimes plays $S_a$ and the latter sometimes plays $s_b$. But then, playing $s_b$ gives the latter player a utility of at most $\frac{1}{k}(-3k) + \frac{k-1}{k}3 < 0$, and it would be better off playing some $S_i$ instead. (Contradiction.) It follows that in no pure-strategy BNE, any element of $S$ is ever played.

Now, in our given pure-strategy equilibrium, consider the set of all the $S_i$ that are played by player 1 for some type. Clearly there can be at most $k$ such sets. We claim they cover $S$. For if they do not cover some element $s_j$, the expected utility of playing $s_j$ for player 2 is 3 (because player 1 never plays any element of $S$). But this means that player 2 (who never plays any element of $S$ either) is not playing optimally. (Contradiction.) Hence, there exists a set cover. ∎

If one allows for general mixed strategies, a Bayes-Nash equilibrium always exists [12]. However, the question of how efficiently one can be constructed remains open.

## 5 Pure-strategy Nash equilibria in stochastic (Markov) games

We now shift our attention from single-shot games to games with multiple stages. In each stage, the players get to act and obtain payoffs. There has already been some research into the complexity of playing repeated and sequential games. For example, determining whether a particular automaton is a best response is $\mathcal{NP}$-complete [5]; it is $\mathcal{NP}$-complete to compute a best-response automaton when the automata under consideration are bounded [37]; the question of whether a given player with imperfect recall can guarantee itself a given payoff using pure strategies is $\mathcal{NP}$-complete [20]; and in general, best-responding to an arbitrary strategy can even be noncomputable [51]. In this section, we present, to our knowledge, the first $\mathcal{PSPACE}$-hardness result on the existence of a pure-strategy equilibrium.

A multi-stage game is typically represented as a *stochastic (Markov) game*, where there is an underlying set of states, and the game shifts between these states from stage to stage [45, 46, 12]. At every stage, each player's payoff depends not only on the players' actions, but also on the state. Furthermore, the probability of transitioning to a given state is determined by the current state and the players' current actions. Hardness results for such games cannot be obtained simply by formulating a known hard game such as generalized Go [26] or QSAT [48] as a Markov game, because such a formulation would have to specify an exponential number of states. Even if the number of states is polynomial, one might suspect hardness due to the fact that the strategy spaces are extremely rich. However, in this section we show $\mathcal{PSPACE}$-hardness even in a variant where the strategy spaces are simple (in the sense that the players cannot condition their actions on events in the game).

**Definition 9** *A* stochastic (Markov) game *consists of the following.*

- *A set of players $A$;*
- *A set of states $S$, among which the game transits;*
- *For each player $i$, a set of actions $\Sigma_i$ that can be played in any state;*
- *A transition probability function $p : S \times \Sigma_1 \times \ldots \times \Sigma_{|A|} \times S \to [0, 1]$, where $p(s_1, a_1, \ldots, a_n, s_2)$ gives the probability of the game being in state $s_2$ in the next stage given that the current state of the game is $s_1$ and the players play actions $a_1, \ldots, a_n$;*
- *For each player $i$, a payoff function $u_i : S \times \Sigma_1 \times \ldots \Sigma_{|A|} \to \Re$, where $u_i(s, a_1, \ldots, a_{|A|})$ gives the payoff to player $i$ in state $s$ where the players play actions $a_1, \ldots, a_n$;*
- *A discount factor $\delta$ such that the total utility of agent $i$ is $\sum\limits_{k=0}^{\infty} \delta^k u_i(s^k, a_1^k, \ldots, a_{|A|}^k)$, where $s^k$ is the state of the game at stage $k$ and the players play actions $a_1^k, \ldots, a_n^k$ in stage $k$.*

In general, a player need not always be aware of the current state of the game, the actions the others played in previous stages, or the payoffs that the player has accumulated. In the extreme case, players never find out any of these and are hence playing blindly. We call such a Markov game *invisible*. It is relatively easy to specify a pure strategy in an invisible Markov game, because there is nothing to condition on. Hence, such a strategy is "simply" an infinite sequence of actions (for player $i$, a sequence $\{a_i^k\}$, where it plays action $a_i^k$ in stage $k$, regardless).[3] In spite of this apparent simplicity of the game, we show that determining whether pure-strategy equilibria exist is extremely hard.

**Definition 10 (PURE-STRATEGY-INVISIBLE-MARKOV-NE)** *We are given an invisible Markov game. We are asked whether there exists a Nash equilibrium where all the strategies are pure.*

We show that this problem is $\mathcal{PSPACE}$-hard, by reducing from PERIODIC-SAT, which is known to be $\mathcal{PSPACE}$-complete [36].

**Definition 11 (PERIODIC-SAT)** *We are given a CNF formula $\phi(0)$ over the variables $\{x_1^0 \ldots x_n^0\} \cup \{x_1^1 \ldots x_n^1\}$. Let $\phi(k)$ be the same formula, except that all the superscripts are incremented by $k$. We are asked whether there exists a Boolean assignment to the variables $\bigcup_{k=0,1,\ldots}\{x_1^k \ldots x_n^k\}$ such that $\phi(k)$ is satisfied for every $k = 0, 1, \ldots$.*

**Theorem 3** *PURE-STRATEGY-INVISIBLE-MARKOV-NE is $\mathcal{PSPACE}$-hard, even when the game is symmetric, 2-player, and the transition process is deterministic.*

---

[3]We do not need to worry about issues of credible threats and subgame perfection in this setting, so we can simply use Nash equilibrium as our solution concept [28].

**Proof**: We reduce an arbitrary PERIODIC-SAT instance to the following symmetric 2-player PURE-STRATEGY-INVISIBLE-MARKOV-NE instance. The state space is $S = \{s_i\}_{1 \leq i \leq n} \cup \{t^1_{i,c}\}_{1 < i \leq 2n; c \in C} \cup \{t^2_{i,c}\}_{1 < i \leq 2n; c \in C} \cup \{r\}$, where $C$ is the set of clauses in $\phi(0)$. Furthermore, $\Sigma \equiv \Sigma_1 = \Sigma_2 = \{t, f\} \cup C$. The transition probabilities are as follows:

- $p(s_i, x^1, x^2, s_{i+1(modn)}) = 1$ for $1 < i \leq n$ and all $x^1, x^2 \in \Sigma$;
- $p(s_1, b^1, b^2, s_2) = 1$ for all $b^1, b^2 \in \{t, f\}$;
- $p(s_1, c, b, t^1_{2,c}) = 1$ for all $b \in \{t, f\}$ and $c \in C$;
- $p(s_1, b, c, t^2_{2,c}) = 1$ for all $b \in \{t, f\}$ and $c \in C$;
- $p(s_1, c^1, c^2, r) = 1$ for all $c^1, c^2 \in C$;
- $p(t^j_{i,c}, x^1, x^2, t^j_{i+1,c}) = 1$ for all $1 < i < 2n$, $j \in \{1, 2\}$, $c \in C$, and $x^1, x^2 \in \Sigma$;
- $p(t^j_{2n,c}, x^1, x^2, r) = 1$ for all $j \in \{1, 2\}$, $c \in C$, and $x^1, x^2 \in \Sigma$;
- $p(r, x^1, x^2, r) = 1$ for all $x^1, x^2 \in \Sigma$.

Some of the utilities obtained in a given stage are as follows (we do not specify utilities irrelevant to our analysis):

- $u_1(s_i, x^1, x^2) = u_2(s_i, x^2, x^1) = 0$ for $1 < i \leq n$ and all $x^1, x^2 \in \Sigma$;
- $u_1(s_1, b^1, b^2) = u_2(s_1, b^2, b^1) = 0$ for all $b^1, b^2 \in \{t, f\}$;
- $u_1(s_1, c, b) = u_2(s_1, b, c) = 1$ for all $b \in \{t, f\}$ and $c \in C$, when setting variable $x^0_1$ to $b$ does not satisfy $c$;
- $u_1(s_1, c, b) = u_2(s_1, b, c) = -1$ for all $b \in \{t, f\}$ and $c \in C$, when setting variable $x^0_1$ to $b$ does satisfy $c$;
- $u_1(s_1, c^1, c^2) = u_2(s_1, c^2, c^1) = -1$ for all $c^1, c^2 \in C$;
- $u_1(t^1_{kn+i,c}, x, b) = u_2(t^2_{kn+i,c}, b, x) = 0$ for $k \in \{0, 1\}$, $1 \leq i \leq n$, all $c \in C$ and $b \in \{t, f\}$ such that setting variable $x^k_i$ to $b$ does not satisfy $c$, and all $x \in \Sigma$;
- $u_1(t^1_{kn+i,c}, x, b) = u_2(t^2_{kn+i,c}, b, x) = -4$ for $k \in \{0, 1\}$, $1 \leq i \leq n$, all $c \in C$ and $b \in \{t, f\}$ such that setting variable $x^k_i$ to $b$ does satisfy $c$, and all $x \in \Sigma$;
- $u_1(t^1_{kn+i,c}, x, c') = u_2(t^2_{kn+i,c}, c', x) = 0$ for $k \in \{0, 1\}$, $1 \leq i \leq n$, all $c, c' \in C$, and all $x \in \Sigma$.

Additionally, the game played in state $r$ is some symmetric zero-sum game without a pure-strategy equilibrium (for example, a generalization of rock-paper-scissors) with very small payoffs. Finally, the discount factor is $\delta = (\frac{1}{2})^{\frac{1}{2n+1}}$ (so that $\delta^{2n} > \frac{1}{2}$).

We start our analysis with a few observations. First, there can be no pure-strategy equilibrium in which state $r$ is reached at some point, because (since $r$ is an absorbing state) this would require that some pure-strategy equilibrium of the game in state $r$ were played whenever state $r$ occurred. (Otherwise a player who is not best-responding in one of these stages could simply switch to a best response in this stage, and because the game is invisible, the rest of the game would remain unaffected, so this would give higher utility.) But such an equilibrium does not exist. Second, if we ever reach one of the $t^j_{i,c}$ states, we will inevitably reach state $r$ at some point after this. It follows that all pure-strategy Nash equilibria never leave the $s_i$ states.

Now suppose an assignment satisfying the periodic SAT formula exists. Let both players play as follows: in stage $kn + i$ (with $1 \leq i \leq n$), $b \in \{t, f\}$ is played, where $b$ is the value that the variable $x_i^k$ is set to. Clearly, both players receive utility 0 with these strategies. Does either player have an incentive to deviate? The only deviation of any significance is to play some $c \in C$ when the current state is $s_1$. So, without loss of generality (because of the symmetry of the game), say player 2 deviates to playing $c \in C$ in stage $kn + 1$ (when the state is $s_1$). We know that in the satisfying assignment, some variable $x_i^l$ among $x_1^k, \ldots, x_n^k, x_1^{k+1}, \ldots, x_n^{k+1}$ is set to some $b$ such that setting $x_i^{l-k}$ to $b$ satisfies $c$. If it is $x_1^k$, which is set to $b$, then in stage $kn + 1$ player 1 plays $b$, and player 2 gets payoff $-1$ in this stage since we are in state $s_1$ and setting $x_1^0$ to $b$ satisfies $c$. Otherwise, if it is $x_i^l$ with $l = k + 1$ or $i \neq 1$, which is set to $b$, then player 2 will get payoff 1 in stage $kn + 1$, but in stage $ln + i$ player 1 plays $b$, and player 2 gets payoff $-4$ in this stage since we are in state $t^2_{(l-k)n+i,c}$ and setting $x_i^{l-k}$ to $b$ satisfies $c$. The discounting is insignificant enough that this more than cancels out the 1 earned in stage $kn+1$. Player 2 will get (at most) 0 in the other stages up to the first stage in state $r$, and given that we made the payoffs in the game in state $r$ sufficiently small relative to $\delta$, player 2 will not earn enough in the remaining stages to cancel out its losses so far. So there is no incentive to deviate. Thus, a pure-strategy NE exists.

On the other hand, suppose that no assignment satisfying the periodic SAT formula exists. Let us investigate whether a Nash equilibrium could exist. We know that in such a Nash equilibrium we never leave the $s_i$, so both players receive utility 0, and no $c$ is ever played in a stage with state $s_1$. Since playing a $c$ in one of the other stages can have no deterrent value, we may suppose that only elements of $\{t, f\}$ are played. Now consider the following assignment to the $x_i^k$: if player 1 plays $b$ in stage $kn + i$, $x_i^k$ is set to $b$. Since no assignment satisfying the periodic SAT formula exists, we know there is some clause $c$ and some $k$ such that no variable $x_i^l$ among $x_1^k, \ldots, x_n^k, x_1^{k+1}, \ldots, x_n^{k+1}$ is set to some $b$ such that setting $x_i^{l-k}$ to $b$ satisfies $c$. But then, if player 2 deviates to play this $c$ in stage $kn+1$, it will receive payoff 1 in this stage, and payoff 0 in all the remaining stages up to the first stage in state $r$. Furthermore, player 2 can guarantee itself at least payoff 0 in each stage in state $r$, as this state corresponds to a zero-sum symmetric game. It follows that this deviation gives player 2 positive utility and is hence beneficial. Thus, no pure-strategy NE exists. ∎

A simpler version of the same argument shows a weaker form of hardness for the case where the game is restricted to have only finitely many stages (we omit the proof due to limited space):

**Theorem 4** *PURE-STRATEGY-INVISIBLE-MARKOV-NE is $\mathcal{NP}$-hard, even when the game is symmetric, 2-player, the transition process is deterministic, and the number of stages in the game is finite.*

# 6 Conclusions and future research

Noncooperative game theory provides a normative framework for analyzing strategic interactions. However, for the toolbox to be operational, the solutions it defines will have to be *computed*.

In this paper, we provided a single reduction that 1) demonstrates $\mathcal{NP}$-hardness of determining whether Nash equilibria with certain natural properties exist, and 2) demonstrates the #$\mathcal{P}$-hardness of counting Nash equilibria (or connected sets of Nash equilibria). We also showed that 3) determining whether a pure-strategy Bayes-Nash equilibrium exists is $\mathcal{NP}$-hard, and that 4) determining whether a pure-strategy Nash equilibrium exists in a stochastic (Markov) game is $\mathcal{PSPACE}$-hard even in invisible games (and remains $\mathcal{NP}$-hard if the game is finite). All of our hardness results hold even if there are only two players and the game is symmetric.

The intersection of computer science and economics is a fertile research area, and has many exciting avenues for future research. For example, within the area of computational complexity, there are open questions regarding the complexity of *executing* various mechanism optimally (e.g., [4, 42, 43, 16]) or approximately (e.g. [33, 25]), the complexity of *manipulating* various mechanisms (e.g., [3, 2, 9, 44]), the complexity of *designing* mechanisms (that lead to desirable outcomes) [10, 40], and the complexity of deciding what information to *elicit* from the players in various mechanisms [11]. Another avenue involves studying more sophisticated equilibrium notions which take into account that players have limited memory (e.g. [32, 41, 1, 14, 39]) or limited capability to solve optimization problems (e.g. [23, 24, 19, 34]). There are also open issues on communication complexity in games (e.g., [35, 7, 8, 17, 50]), and on the complexity of computing general equilibria ("market equilibria") (e.g., [50]) and other solutions.

There are numerous open research questions even in the area of computing solutions to noncooperative games. Some recent work has focused on novel knowledge representations which, in certain settings, can drastically speed up equilibrium finding (e.g. [21, 18, 22]). One avenue of future theoretical work includes identifying restricted classes of games for which equilibria (or equilibria with certain properties) can be found fast. Another avenue involves studying the complexity of characterizing (some of) the equilibria of a game *partially*. Yet another avenue includes analyzing the computational complexity of other solution concepts from noncooperative game theory.

# References

[1] Dilip Abreu and Ariel Rubinstein. The structure of Nash equilibrium in repeated games with finite automata. *Econometrica*, 56(6):1259–1281, 1988.

[2] John J. Bartholdi, III and James B. Orlin. Single transferable vote resists strategic voting. *Social Choice and Welfare*, 8(4):341–354, 1991.

[3] John J. Bartholdi, III, Craig A. Tovey, and Michael A. Trick. The computational difficulty of manipulating an election. *Social Choice and Welfare*, 6(3):227–241, 1989.

[4] John J. Bartholdi, III, Craig A. Tovey, and Michael A. Trick. Voting schemes for which it can be difficult to tell who won the election. *Social Choice and Welfare*, 6:157–165, 1989.

[5] E. Ben-Porath. The complexity of computing a best response automaton in repeated games with mixed strategies. *Games and Economic Behavior*, pages 1–12, 1990.

[6] Francis Chu and Joseph Halpern. On the NP-completeness of finding an optimal strategy in games with common payoffs. *International Journal of Game Theory*. To appear.

[7] Wolfram Conen and Tuomas Sandholm. Preference elicitation in combinatorial auctions: Extended abstract. In *Proceedings of the ACM Conference on Electronic Commerce (ACM-EC)*, pages 256–259, Tampa, FL, October 2001. A more detailed description of the algorithmic aspects appeared in the IJCAI-2001 Workshop on Economic Agents, Models, and Mechanisms, pp. 71–80.

[8] Wolfram Conen and Tuomas Sandholm. Partial-revelation VCG mechanism for combinatorial auctions. In *Proceedings of the National Conference on Artificial Intelligence (AAAI)*, Edmonton, Canada, 2002.

[9] Vincent Conitzer and Tuomas Sandholm. Complexity of manipulating elections with few candidates. In *Proceedings of the National Conference on Artificial Intelligence (AAAI)*, Edmonton, Canada, 2002.

[10] Vincent Conitzer and Tuomas Sandholm. Complexity of mechanism design. 2002. Under review.

[11] Vincent Conitzer and Tuomas Sandholm. Vote elicitation: Complexity and strategy-proofness. In *Proceedings of the National Conference on Artificial Intelligence (AAAI)*, Edmonton, Canada, 2002.

[12] Drew Fudenberg and Jean Tirole. *Game Theory*. MIT Press, 1991.

[13] I. Gilboa and E. Zemel. Nash and correlated equilibria: Some complexity considerations. *Games and Economic Behavior*, 1989.

[14] Itzhak Gilboa and Dov Samet. Bounded versus unbounded rationality: The tyranny of the weak. *Games and Economic Behavior*, pages 213–221, 1989.

[15] John Harsanyi. Game with incomplete information played by Bayesian players. *Management Science*, 14:159–182; 320–334; 486–502, 1967–68.

[16] John Hershberger and Subhash Suri. Vickrey prices and shortest paths: What is an edge worth? In *Proceedings of the Symposium on Foundations of Computer Science (FOCS)*, 2001.

[17] Benoit Hudson and Tuomas Sandholm. Effectiveness of preference elicitation in combinatorial auctions. Technical report, Carnegie Mellon Computer Science, CMU-CS-02-124, March, 2002. Also accepted to the Stanford Institute for Theoretical Economics workshop (SITE), June, 2002.

[18] Michael Kearns, Michael Littman, and Satinder Singh. Graphical models for game theory. In *Proceedings of the Uncertainty in Artificial Intelligence Conference (UAI)*, 2001.

[19] Noa Kfir-Dahav, Dov Monderer, and Moshe Tennenholtz. Mechanism design for resource bounded agents. In *Proceedings of the Fourth International Conference on Multi-Agent Systems (ICMAS)*, 2000.

[20] D. Koller and N. Megiddo. The complexity of twoperson zero-sum games in extensive form. *Games and Economic Behavior*, pages 4:528–552, 1992.

[21] Daphne Koller, Nimrod Megiddo, and Bernhard von Stengel. Efficient computation of equilibria for extensive two-person games. *Games and Economic Behavior*, 14(2):247–259, 1996.

[22] Daphne Koller and B Milch. Multi-agent influence diagrams for representing and solving games. In *Proceedings of the Seventeenth International Joint Conference on Artificial Intelligence (IJCAI)*, pages 1027–1034, Seattle, WA, 2001.

[23] Kate Larson and Tuomas Sandholm. Bargaining with limited computation: Deliberation equilibrium. *Artificial Intelligence*, 132(2):183–217, 2001. Short early version appeared in the Proceedings of the National Conference on Artificial Intelligence (AAAI), pp. 48–55, Austin, TX, 2000.

[24] Kate Larson and Tuomas Sandholm. Costly valuation computation in auctions. In *Theoretical Aspects of Rationality and Knowledge (TARK VIII)*, pages 169–182, Sienna, Italy, July 2001.

[25] Daniel Lehmann, Lidian Ita O'Callaghan, and Yoav Shoham. Truth revelation in rapid, approximately efficient combinatorial auctions. In *Proceedings of the ACM Conference on Electronic Commerce (ACM-EC)*, pages 96–102, Denver, CO, November 1999.

[26] D Lichtenstein and M Sipser. GO is polynomial-space hard. *Journal of the ACM*, 27:393–401, 1980.

[27] R. Duncan Luce and Howard Raiffa. *Games and Decisions*. John Wiley and Sons, New York, 1957. Dover republication 1989.

[28] Andreu Mas-Colell, Michael Whinston, and Jerry R. Green. *Microeconomic Theory*. Oxford University Press, 1995.

[29] Andrew McLennan. On the expected number of nash equilibria of a normal form game. 1997.

[30] Andrew McLennan and In-Uck Park. Generic 4x4 two person games have at most 15 nash equilibria. 1997.

[31] John Nash. Equilibrium points in n-person games. *Proc. of the National Academy of Sciences*, 36:48–49, 1950.

[32] A Neyman. Bounded complexity justifies cooperation in finitely repeated prisoner's dilemma. *Economic Letters*, pages 227–229, 1985.

[33] Noam Nisan and Amir Ronen. Algorithmic mechanism design. In *Proceedings of the Annual ACM Symposium on the Theory of Computing (STOC)*, pages 129–140, 1999.

[34] Noam Nisan and Amir Ronen. Computationally feasible VCG mechanisms. In *Proceedings of the ACM Conference on Electronic Commerce (ACM-EC)*, pages 242–252, Minneapolis, MN, 2000.

[35] Noam Nisan and Ilya Segal. The communication complexity of efficient allocation problems, 2002. Draft. Second version March 5th.

[36] J Orlin. The complexity of dynamic languages and dynamic optimization problems. In *Proceedings of the Annual ACM Symposium on the Theory of Computing (STOC)*, pages 218–227, 1981.

[37] C. H. Papadimitriou. On players with a bounded number of states. *Games and Economic Behavior*, pages 122–131, 1992.

[38] Christos Papadimitriou. Algorithms, games and the Internet. In *Proceedings of the 33rd Annual ACM Symposium on the Theory of Computing*, pages 749–253, 2001.

[39] Christos H. Papadimitriou and Mihalis Yannakakis. On complexity as bounded rationality. In *Proceedings of the Annual ACM Symposium on the Theory of Computing (STOC)*, pages 726–733, 1994.

[40] Tim Roughgarden and Eva Tardos. Designing networks for selfish users is hard. In *Proceedings of the Symposium on Foundations of Computer Science (FOCS)*, 2001.

[41] Ariel Rubinstein. Finite automata play the repeated prisoner's dilemma. *Journal of Economic Theory*, 39:83–96, 1986.

[42] Tuomas Sandholm. Algorithm for optimal winner determination in combinatorial auctions. *Artificial Intelligence*, 135:1–54, January 2002. First appeared as an invited talk at the First International Conference on Information and Computation Economies, Charleston, SC, Oct. 25–28, 1998. Extended version appeared as Washington Univ.,

Dept. of Computer Science, tech report WUCS-99-01, January 28th, 1999. Conference version appeared at the International Joint Conference on Artificial Intelligence (IJCAI), pp. 542–547, Stockholm, Sweden, 1999.

[43] Tuomas Sandholm and Subhash Suri. Market clearability. In *Proceedings of the Seventeenth International Joint Conference on Artificial Intelligence (IJCAI)*, pages 1145–1151, Seattle, WA, 2001.

[44] Tuomas Sandholm and Fredrik Ygge. Constructing speculative demand functions in equilibrium markets. Technical Report WUCS-99-26, Washington University, Department of Computer Science, October 1999. Short early version appeared at the International Joint Conference on Artificial Intelligence (IJCAI), pp. 632–638, Nagoya, Japan, 1997.

[45] Lloyd S Shapley. Stochastic games. *Proc. of the National Academy of Sciences*, 39:1095–1100, 1953.

[46] M Sobel. Noncooperative stochastic games. *Annals of Mathematical Statistics*, 42:1930–1935, 1971.

[47] Bernhard von Stengel. Computing equilibria for two-person games. In Robert Aumann and Sergiu Hart, editors, *Handbook of game theory*, volume 3. North Holland, Amsterdam, 2002. Forthcoming.

[48] L Stockmeyer and A Meyer. Word problems requiring exponential time. In *Proceedings of the Annual ACM Symposium on the Theory of Computing (STOC)*, pages 1–9, 1973.

[49] Leslie Valiant. The complexity of computing the permanent. *Theoretical Computer Science*, 8:189–201, 1979.

[50] Deng Xiaotie, Christos Papadimitriou, and Muli Safra. On the complexity of equilibria. In *Proceedings of the Annual ACM Symposium on the Theory of Computing (STOC)*, 2002.

[51] William R. Zame and John H. Nachbar. Non-computable strategies and discounted repeated games. *Economic Theory*, pages (1)103–122, 1996.